\documentclass{PoS}

\usepackage{newtxtext,newtxmath}
\usepackage[T1]{fontenc}
\usepackage{ae,aecompl}
\usepackage{graphicx}	% Including figure files
\usepackage{amsmath}	% Advanced maths commands
\usepackage{amssymb}	% Extra maths symbols

\title{Simultaneous Spectral and Spatial Modelling of Young Pulsar Wind Nebulae}

\ShortTitle{Spectral and Spatial Modelling of PWNe}

\author{\speaker{Carlo van Rensburg}, Christo Venter, and P. Paulus Kr\"uger\\
        Centre for Space Research, North-West University, Potchefstroom Campus, Private Bag X6001, Potchefstroom, South Africa, 2520\\
        E-mail: \email{carlo.rensburg@gmail.com}}

\abstract
{
We model the morphology and spectrum of a pulsar wind nebula using a leptonic emission code. This code is a time-dependent, multi-zone model that investigates the changes in the particle spectrum as they traverse the nebula. We calculate the synchrotron and inverse Compton emissivities at different positions in the nebula, obtaining the surface brightness versus the radius, and also the size of the nebula as a function of energy. We incorporate a time and spatially-dependent $B$-field, spatially-dependent bulk particle speed implying convection and adiabatic losses, diffusion, as well as radiative losses.  We calibrate our new model using two independent models. We then apply the model to PWN G0.9+0.1 and show that simultaneously fitting the spectral energy distribution and the energy-dependent source size may lead to constraints on several model parameters pertaining to the spatial properties of the PWN.}

\FullConference{5th Annual Conference on High Energy Astrophysics in Southern Africa\\
		4-6 October, 2017\\
		University of the Witwatersrand (Wits), South Africa}

\begin{document}

\section{Introduction}
Discoveries during the last decade by ground-based Imaging Atmospheric Cherenkov Telescopes (IACTs) have increased the number of known very-high-energy (VHE, $E >$ 100~GeV) $\gamma$-ray sources to nearly 200\footnote{http://tevcat2.uchicago.edu/} Nearly 40 of these are confirmed pulsar wind nebulae (PWNe) \cite{Hewitt2015}. Following the nine-year H.E.S.S. Galactic Plane Survey (HGPS; \cite{HGPS2016}), H.E.S.S.\ published a paper describing the properties of 19 PWNe and 10 strong PWN candidates, as well as empirical trends between several PWN/pulsar parameters \cite{PWN_pop2017}. It is expected that the future Cherenkov Telescope Array (CTA), with its order-of-magnitude increase in sensitivity and improvement in angular resolution, will discover several more (older and fainter) PWNe and reveal many more morphological details. A systematic search with the \textit{Fermi} Large Area Telescope (LAT) for GeV emission in the vicinity of TeV-detected sources yielded 5 high-energy $\gamma$-ray PWNe and eleven PWN candidates \cite{3FGL2015}. In the X-ray to VHE $\gamma$-ray energy range there are 85 PWNe or PWN candidates with 71 of them having associated pulsars \cite{Kargaltsev2012}. PWNe are also observed at lower energies. If the pulsar is very slowly moving one might observe a composite supernova remnant (SNR), with nebular and shell emission visible in both radio and X-ray bands. Such young systems exhibit a high degree of spherical symmetry and it is possible that the SNR reverse shock has not yet interacted with the PWN (e.g., SNR G11.2$-$0.3 and G21.5$-$0.9). The PWN around PSR B1509$-$58 provides a counter example, exhibiting a strong anti-correlation between the radio and X-ray emission morphology. This system is reminiscent of older PWNe associated with fast-moving pulsars and $\gamma$-ray sources that exhibit complex morphologies (e.g., the Rabbit Nebula and G327.1$-$1.1) \cite{Roberts2005, SlanePWN2017}. In older PWN, a decreasing $B$-field may lead to $\gamma$-ray emission dominating the observed radio and X-ray emission (e.g., HESS J1825$-$137) \cite{SlanePWN2017}. High-resolution observations by \textit{Chandra} X-ray Observatory have furthermore revealed complex substructures such as toroidal structures, bipolar jets, and filaments \cite{Helfand2001, Roberts2003}. Similarly, high-resolution radio images sometimes reveal complex PWN morphology including filaments, knots, and holes \cite{Dubner2008}. Complementary optical and infrared observations of PWN may uncover spectral features in the PWN particle spectrum, and information about the shocked supernova ejecta, and newly formed dust~\cite{Slane2017_T}. 

Currently there are two main avenues of modelling PWNe: magnetohydrodynamic (MHD) and leptonic codes. MHD models can model the morphology of PWNe in great detail \cite{Bucciantini2014}, but they can not predict their spectral energy distributions (SEDs). These models describe the geometry and environments of the PWN and not the high-energy particle spectrum responsible for the broadband emission. Therefore, the information about the radiation spectrum is lost. On the other hand, leptonic models excel in predicting the PWN SEDs, but most of these model are single-sphere models and are thus unable to reproduce any of the morphological characteristics of these PWNe.

In light of the above, there is a void in the current modeling landscape for a spatio-temporal and energy-dependent PWN model that calculates both the morphology and the SED of a PWN. By adding a spatial dimension to an emission code, one is able to constrain the model more significantly using available data such as surface brightness profiles, spectral index versus radius, and energy-dependent source size, and thus probe the PWN physics more deeply. Our newly developed time-dependent, multi-zone model aids in breaking model parameter degeneracies by first constraining the profiles of, e.g., the PWN $B$-field and then fitting the observed SED in a more constrained parameter space, thus making use of both spectral and spatial data. Development of such a model would place one in a position to interpret the anticipated morphological details that will be measured by future experiments, e.g., CTA. We implemented such a model \cite{vRensburg2014, CvR2015} and discuss some of the results here. For more details, see Van Rensburg et al., submitted. 

\section{The Model}
We developed and implemented a multi-zone, time-dependent code, which models the transport of particles through a PWN. We make the simplifying assumption that the geometrical structure of the PWN may be modelled as a sphere into which particles are injected and allowed to diffuse and undergo energy losses. Thus, the particle transport is treated as being spherically symmetric and the only changes in the particle spectrum will be in the radial direction (apart from changes in the particle energy). The model therefore consists of three dimensions in which the transport equation is solved: the spatial or radial dimension, the lepton energy dimension, and the time dimension. 

We solve the following equation
% \begin{equation}
% \begin{split}
% \frac{\partial N_{\rm{e}}}{\partial t} =& -\mathbf{V} \cdot (\nabla N_{\rm{e}}) +  \kappa \nabla^2 N_{\rm{e}}  \\
% &+ \frac{1}{3}(\nabla \cdot \mathbf{V})\left( \left[\frac{\partial N_{\rm{e}}}{\partial \ln E_{\rm{e}}} \right] - 2N_{\rm{e}} \right)   \\
% &+ \frac{\partial }{\partial E}(\dot{E}_{\rm{e,tot}}N_{\rm{e}}) +  Q(\mathbf{r},E_{\rm{e}},t),
% \end{split}
% \label{eq:transportFIN}
% \end{equation} 

\begin{equation}
\begin{split}
\frac{\partial N_{\rm{e}}}{\partial t} =& -\mathbf{V} \cdot (\nabla N_{\rm{e}}) +  \kappa \nabla^2 N_{\rm{e}} + \frac{1}{3}(\nabla \cdot \mathbf{V})\left( \left[\frac{\partial N_{\rm{e}}}{\partial \ln E_{\rm{e}}} \right] - 2N_{\rm{e}} \right) + \frac{\partial }{\partial E}(\dot{E}_{\rm{e,tot}}N_{\rm{e}}) +  Q(\mathbf{r},E_{\rm{e}},t),
\end{split}
\label{eq:transportFIN}
\end{equation} 

with $N_{\rm{e}}$ the number of particles per unit energy and volume, $\mathbf{V}$ the bulk motion of particles, $\kappa$ the energy-dependent diffusion coefficient, $\dot{E}_{\rm{e,tot}}$ the total energy loss rate, including radiation and adiabatic energy losses, and $Q$ the injection spectrum. We model the particle injection spectrum using a broken power law \cite{VdeJager2007} and use the pulsar's spin-down luminosity to normalise this spectrum. We consider synchrotron radiation (SR) and inverse Compton (IC) scattering, similar to calculations done by Kopp et al.~\cite{Kopp2013} in their globular cluster model, as well as adiabatic energy losses due to the bulk motion of the particles. As a first approximation, we implement Bohm-type diffusion, $\kappa(E_{\rm e}) = \kappa_0(E_{\rm e}/E'_{0})$ with $E'_{0}$ = 1~TeV.

We parametrise the bulk particle speed inside the PWN as 
\begin{equation}
V(r) = V_0\left(\frac{r}{r_0}\right)^{\alpha_{\rm{V}}},
\label{V_profile}
\end{equation}
with $\alpha_{\rm{V}}$ the velocity profile parameter. We also use a parametrised form of the $B$-field
\begin{equation}
B(r,t) = B_{\rm{age}}\left(\frac{r}{r_0}\right)^{\alpha_{\rm{B}}}\left(\frac{t}{t_{\rm{age}}}\right)^{\beta_{\rm{B}}},
\label{B_Field}
\end{equation}
with $B_{\rm{age}}$ the present-day $B$-field at $r = r_0$ and $t = t_{\rm{age}}$, $t$ the time since the PWN's birth, $t_{\rm{age}}$ the PWN age, and $\alpha_{\rm{B}}$ and $\beta_{\rm{B}}$ the $B$-field parameters. The $B$-field and bulk motion are linked by Faraday's law of induction \cite{Ferreira_deJager2008}
\begin{equation}
\frac{\partial \mathbf{B}}{\partial t} = \nabla \times \left(\mathbf{V} \times \mathbf{B}  \right).
\label{eq:B_V}
\end{equation}
By assuming that the temporal change in the magnetic field is slow, we can set $\partial B/\partial t \sim 0$. Also assuming that the system is spherically symmetric, one can show that \cite{Kennel1984} $VBr={\rm{constant}}=V_0B_0r_0,$
% \begin{equation}
% VBr={\rm{constant}}=V_0B_0r_0,
% \label{eq:vbr=c}
% \end{equation}
and thus we have a relation between the radial component of the particle bulk motion and the radial component of the $B$-field\footnote{This relation follows from the parametrised form of the $B$-field and the bulk motion.}: $\alpha_{\rm{V}}+\alpha_{\rm{B}}=-1.$
% \begin{equation}
% \alpha_{\rm{V}}+\alpha_{\rm{B}}=-1.
% \label{eq:a_v+a_b=-1}
% \end{equation}  

We implement a reflective inner boundary\footnote{We implement a reflective inner boundary by requiring zero particle flux at the innermost radial bin. We apply this to all particles independent of energy or pitch angle.} to avoid losing particles towards the pulsar past the termination shock, and an escape outer boundary $r_{\rm{max}}$. We limit the maximum particles' energy \cite{VdeJager2007} by requiring that 
\begin{equation}
E_{\rm{max}} \leq \frac{e}{2}\sqrt{\frac{L(t)\sigma}{c(1+\sigma)}}.
\label{eq:E_max_par}
\end{equation}
This condition follows from a containment argument. For our IC calculation, we chose three target photon fields, i.e., the cosmic background radiation (CMB), Galactic background infrared (IR) photons, and starlight. 

\section{Studying the model behaviour}

We performed a line-of-sight (LOS) calculation to project the radiation onto the plane of the sky in order to find the surface brightness and flux as a function of 2D projected radius. This allows us to estimate the size of the PWN and also study its size as a function of energy. 

We calibrated our code by comparing it, with the same set of parameters, to two previous models \cite{VdeJager2007, Torres2014}. Our predicted SED differed by $\sim20\%$ and $\sim50\%$, respectively.

We preformed a thorough parameter study to investigate the effects that all the different parameters have on the code. For the full parameter study see Van Rensburg \cite{CvR2015}. Here we show the effects that changing two spatial parameters ($\kappa_0$ and $V_0$) and keeping the others fixed has on the code. 

\begin{figure}[t]
\includegraphics[width=0.9\textwidth]{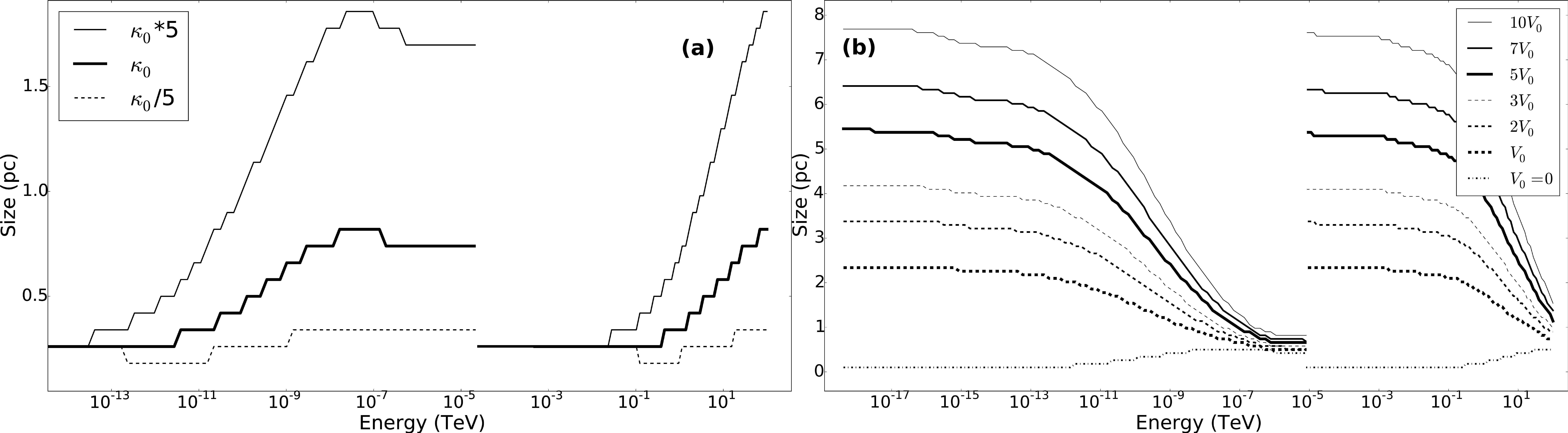}
\caption{\label{fig:SvE_k}\textit{Panel (a):} Size of the PWN as a function of energy when the normalisation constant of the diffusion coefficient is changed. \textit{Panel (b):} Size of the PWN as a function of energy for different normalisations of the bulk particle speed.}
\end{figure}

% \begin{figure}
% \includegraphics[width=.6\textwidth]{SvE_Carlo_V.pdf}
% \caption{\label{fig:SvE_V_Carlo}Size of the PWN as a function of energy for different normalisations of the bulk particle speed.}
% \end{figure}

Figure~\ref{fig:SvE_k}(a) shows how the size of the PWN changes with energy for three different scenarios. The thin solid lines indicate 5$\kappa_0$, the thick solid lines indicate $\kappa_0$, and the dashed lines indicate $\kappa_0 /$5, with $\kappa_0 = cE'_0/3eB$. In this paper for simplicity we assume that $B$ is constant throughout the PWN, i.e., $\alpha_{\rm B} = 0$\footnote{In future we will treat $\alpha_{\rm B}$ as a free parameter to allow us to more accurately fit the observables, e.g., surface brightness profile and photon index vs.\ $r$.}. For this set of scenarios the size of the PWN increases with increased energy. In the first two scenarios, diffusion dominates the particle transport and causes the high-energy particles to diffuse outward faster than low-energy particles, filling up the outer zones and resulting in a larger size for the PWN at high energies. This effect is larger for high-energy particles due to the energy dependence of the diffusion coefficient ($\kappa \propto E_{\rm e}^{1.0}$). For $\kappa_0/$5, we see that the effect is not as pronounced. Here the diffusion coefficient is so small that the SR energy loss timescale starts to dominate the diffusion timescale. The particles therefore ``burn off" or expend their energy before they can reach the outer zones. In Figure~\ref{fig:SvE_k}(b) the PWN size increases monotonically with $V_0$. At the lower energies convection dominates the radiative energy losses and therefore the particles have a very long lifetime, allowing them to diffuse to the outer zones and still be able to radiate, resulting in a large source size. In contrast to this, at high energies, the SR losses dominate convection, resulting in a very short lifetime for the high-energy particles. Therefore these particles radiate all their energy before they have time to convect to the outer zones. This leads to a relatively smaller X-ray source size.

\section{Concurrent modelling of the SED and energy-dependent PWN size}
\begin{figure}
\includegraphics[width=0.9\textwidth]{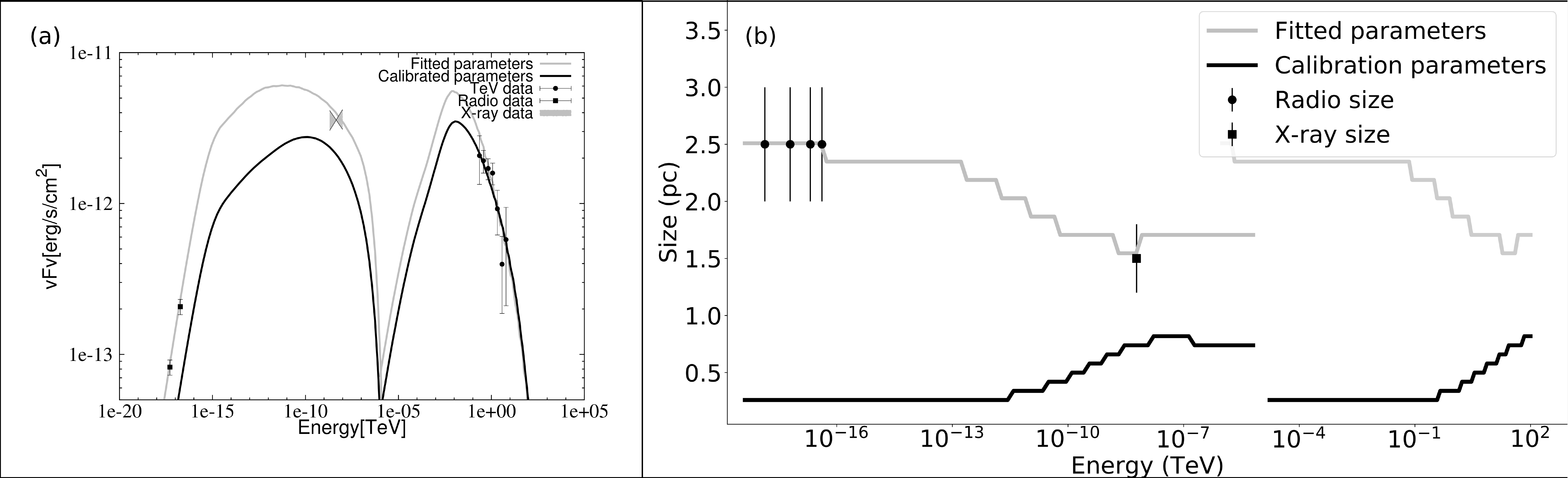}
\caption{\label{fig:SED_combine}\textit{Panel (a):} SED for PWN G0.9+0.1 for the fitted parameters as in Table \ref{tbl:J1356_Carlo}. The radio \cite{HelfandB1987}, X-ray~\cite{Porquet2003} and $\gamma$-ray data \cite{G0.9+0.1_HESS} are also shown. \textit{Panel (b):} Size of the PWN as a function of energy for the fitted parameters in Table~\ref{tbl:J1356_Carlo}. The observed radio \cite{Dubner2008} and X-ray sizes \cite{Porquet2003} are also indicated.}
\end{figure}

Figure~\ref{fig:SED_combine}(a) shows the radiation spectrum for PWN G0.9+0.1 for the parameters as modelled by Torres et al.~\cite{Torres2014} and Figure~\ref{fig:SED_combine}(b) shows our size versus energy model output, assuming these same parameters (black lines), with the dots indicating the radio and the square the X-ray sizes. The upper limit on the predicted TeV size is 10.4 pc, i.e., we use the point spread function of the H.E.S.S. telescope (not shown). The radio data are from Helfand et al. \cite{HelfandB1987} and Dubner et al. \cite{Dubner2008}, the X-ray data are from Porquet et al. \cite{Porquet2003}, and the TeV data from  Aharonian et al. \cite{G0.9+0.1_HESS}. The model provides reasonable fits to the spectral radio, X-ray, and TeV data, however, it is clear that the predicted size of the PWN does not fit the data at all. After adjusting some parameters, we found a better fit and this can also be seen in Figure \ref{fig:SED_combine} (grey lines). Table \ref{tbl:J1356_Carlo} shows the new parameters used for this fit.

% \begin{figure}
% \includegraphics[width=.6\textwidth]{SvE_combine.pdf}
% \caption{\label{fig:SvE_combine}\textit{Panel (a):}Size of the PWN as a function of energy for the fitted parameters in Table~\ref{tbl:J1356_Carlo}. The observed radio \cite{Dubner2008} and X-ray sizes \cite{Porquet2003} are also indicated.}
% \end{figure}

\begin{table}
\begin{center}
\caption{Modified parameters for PWN G0.9+0.1 for concurrently fitting the SED as well as the energy-dependent size of the PWN.\label{tbl:J1356_Carlo}}
\resizebox{.6\textwidth}{!}{
\begin{tabular}{crrr}
\hline
Model Parameter & Symbol & This work & Torres et al. \cite{Torres2014}\\
\hline
Present-day $B$-field & $B(t_{\rm{age}})$ & 16 $\mu \rm{G}$ & 14.0 $\mu \rm{G}$\\
Age of the PWN & $t_{\rm{age}}$ & $3~227~\rm{yr}$ &  $2~000~\rm{yr}$ \\
Initial spin-down power ($10^{38}\rm{erg}$ $\rm{s^{-1}}$)& $L_0$ & 1.44  & 1.0\\ 
$B$-field parameter & $\alpha_{\rm{B}}$ & 0.0 & 0\\
$B$-field parameter & $\beta_{\rm{B}}$ & $-$1.0 & $-1.3$\\
$V$ parameter & $\alpha_{\rm{V}}$ & $-1.0$ & $1.0$\\
Particle bulk motion & $V_0$ & $0.0615~c$ & $1.63 \times 10^{-4} c$\\
Diffusion & $\kappa_0$ & $3.356$ & $1.0$\\
\hline
\end{tabular}
}
\end{center}
\end{table}

The process of finding a better fit to both the SED of the PWN and the energy-dependent size was facilitated by our prior parameter study. The only way in which we could increase the size of the PWN at lower energies was to increase the bulk speed of the particles. This, however, increased the adiabatic energy losses and resulted in a lower radiation spectrum. We countered this by increasing the age of the PWN (which effectively leads to an increase in $L_0$). The bulk speed of the particles had to be increased substantially to fit the data, but given the large errors on the size of the PWN in the radio band,  we could still fit the data with a bulk speed as small as $0.073c$, with $c$ the speed of light in vacuum. We also changed the profile trends for the $B$-field and the bulk speed of the particles. To increase the size of the PWN further we also increased the normalisation of the diffusion coefficient of the particles. This is not a bad assumption as the diffusion was originally modelled to be Bohm-type diffusion, which is a very slow perpendicular diffusion with respect to the $B$-field lines. All these changes produced the grey lines in Figure \ref{fig:SED_combine}. Here we see that we have a good fit for the radio size, which according to data, does not change with energy and the model reproduces this trend as well as the trend where the size of the PWN decreases with increasing energy. This is a common feature of PWNe where cooling dominates other processes.

In a future paper a more robust statistical method may be used to find the best concurrent fit to this source's SED and energy-dependent size and to also investigate the parameter degeneracy.

\section{Conclusions} % Write in your own chapter title
\label{sec:concl}
We solved a Fokker-Planck-type transport equation to model the particle evolution inside a PWN, injecting a broken power-law particle spectrum and allowing this spectrum to evolve over time, taking into account energy losses due to SR, IC scattering, and adiabatic cooling of the PWN due to expansion. We also took into account particle diffusion and convection in the form of a bulk particle motion. Our model is now able to not only fit the observed radiation spectra the PWN, but also yields results concerning the morphology of the PWN (i.e., it is able to reproduce the size of the PWN as a function of energy). Thus we can potentially derive stronger constraints on key quantities characterising the PWN.

In future we will perform a population study to investigate currently known trends, e.g., the X-ray luminosity that correlates with the pulsar spin-down luminosity and its anti-correlation with the characteristic age of the pulsar. We could also probe other trends, e.g., the correlation between the TeV surface brightness of the PWN and the spin-down luminosity of the pulsar, as well as the possible anti-correlation between the surface brightness and the age \cite{PWN_pop2017}. Furthermore, the code is currently only applicable to young PWNe. This should be addressed so that all ages of PWNe can be modelled, e.g., by including a more complex parametrisation of the $B$-field and adding the effect of an asymmetric reverse shock to the code. Lastly we will investigate the effect of changing the outer boundary condition on the predicted size versus energy behaviour.

The CTA will reveal more sources and more information regarding the morphology of PWNe due to its improved sensitivity and angular resolution. This will necessitate the continued development, application, and refinement of spatially-dependent PWN codes such as the one discussed here.

\section{Acknowledgements}
This work is based on the research supported wholly / in part by the National Research Foundation (NRF) of South Africa (Grant Numbers 87613, 90822, 92860, 93278, and 99072). The Grantholder acknowledges that opinions, findings and conclusions or recommendations expressed in any publication generated by the NRF supported research is that of the author(s), and that the NRF accepts no liability whatsoever in this regard.

%\begin{thebibliography}{99}
\bibliographystyle{JHEP}
\bibliography{Bibliography}

%\end{thebibliography}

\end{document}